# Logarithmic decay in single-particle relaxations of hydrated lysozyme powder

Marco Lagi[1,2], Piero Baglioni[2] and Sow-Hsin Chen[1,*]

[1]*Department of Nuclear Science and Engineering, Massachusetts Institute of Technology, Cambridge, MA 02139*
[2]*Department of Chemistry and CSGI, University of Florence, Florence, I 50019, Italy*

We present the self-dynamics of protein amino acids of hydrated lysozyme powder around the physiological temperature by means of molecular dynamics (MD) simulations. The self-intermediate scattering functions (SISF) of the amino acid residue center-of-mass and of the protein hydrogen atoms display a logarithmic decay over 3 decades of time, from 2 picoseconds to 2 nanoseconds, followed by an exponential $\alpha$-relaxation. This kind of slow dynamics resembles the relaxation scenario within the $\beta$-relaxation time range predicted by the mode coupling theory (MCT) in the vicinity of higher-order singularities. These results suggest a strong analogy between the single-particle dynamics of the protein and the dynamics of colloidal, polymeric and molecular glass-forming liquids.

PACS numbers: 61.20.ja, 64.70.kj, 87.14.E-

It is well known that the dynamics of native globular proteins has much in common with the dynamics of glass forming liquids [1-6]. The reason for such a similarity has to be identified among the essential characteristics of these two types of material. They both consist of non-crystalline packing in which their constituents (either molecules in the case of glassy liquids or amino acid residues in the case of proteins) assemble. They also have a complex *energy landscape*, composed of a large number of alternative conformations at similar energies [1].

The analogy between a protein and a glass-former can be seen from the following similarities: 1) at low temperatures proteins undergo the so-called *glass transition* [2], a sudden change of slope in their mean square displacement as a function of temperature, interpreted as the onset of anharmonic processes; 2) the low-energy inelastic spectra of proteins and their hydration water display a feature known as *boson peak*, typical of strong glass formers [3]; 3) the protein *denaturation* can be seen as a sort of strong-to-fragile liquid transition [4], where the folding heavily decreases the number of liquid-like degrees of freedom; 4) proteins have two types of equilibrium fluctuations, the cooperative $\alpha$ (involving large domains of the biomolecule) and the local $\beta$ (involving side-chains), typical of glass-formers [5]; 5) proteins exhibit both short and intermediate range orders, and the construction of a random elastic network using these structures leads naturally to the physics of a glassy material [6].

Proteins and glasses are complex systems, and one of the distinctive features of complex systems is a slow non-exponential relaxation of the density correlation functions $\phi_q(t)$ and of the tagged-particle correlation functions $\phi_q^S(t)$, observed in a wide range of time scales. The time dependence of the relaxation scenario usually follows these three steps: it begins with (a) a short-time gaussian-like ballistic region, followed by (b) the $\beta$-relaxation region which is governed by either two power-law decays $\phi_q(t) \sim (t/\tau_q^\beta)^{-a}$ and $\phi_q(t) \sim (-t/\tau_q^\beta)^b$ or a logarithmic decay $\phi_q(t) \sim A_q - B_q \ln(t/\tau^\beta)$, which then evolves into (c) an $\alpha$-relaxation region that is governed by a stretched exponential decay (or Kohlrausch-Williams-Watts law), $\phi_q(t) \sim \exp(-t/\tau_q^\alpha)^\beta$. These types of relaxation are characteristic of complex systems [7], just as the simple exponential relaxation (or Debye law) $\phi_q(t) \sim \exp(-t/\tau_q)$ is typical for gases and liquids.

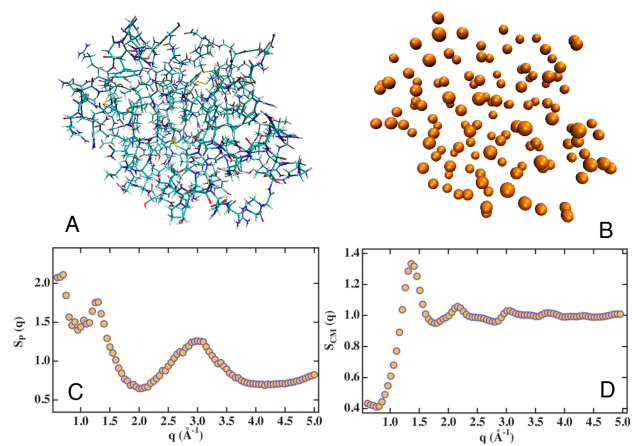

Figure 1. (Color online) Illustration of the analysis of the protein dynamics. A) All-atom representation of lysozyme, B) Visualization of the 129 COM of the lysozyme amino acid residues; C) All-atom static structure factor $S(q)$; D) COM $S(q)$ from MD simulations. $S(q)$ were calculated at T = 300 K, averaged over $1 \cdot 10^3$ configurations and $2 \cdot 10^4$ $q$-directions.

Among these, the *logarithmic decay* is the slowest one and also the least common. In the past years, it has been experimentally found in the time evolution of a wide variety of complex strong interacting systems such as spin glasses [8], granular materials [9], simple glass-forming liquids [10,11], colloidal solutions [12], polymers [13] and protein kinetics [14,15]. To this large number of experimental systems, we can add many numerical simulations on short-ranged attractive colloids [16-18], polymer blends [19], protein folding [20,21] and kinetically constrained models [22].

Starting from 1989, Gotze and collaborators have shown that this particular feature is predicted by the idealized mode coupling theory (MCT) for systems close to a higher-order glass-transition singularity [23-27]. In its ideal version, MCT predicts a sharp transition from an ergodic liquid to a nonergodic arrested state at a critical value $x_c$ of the relevant control parameter $x$ (commonly, the volume fraction or temperature). In the standard MCT formalism, if $n$ is the number of control parameters ($x_1$, $x_2$, ..., $x_n$) the transition is denoted as $A_{n+1}$. Therefore, the standard liquid-glass transition mentioned above is denoted as $A_2$. However, higher order transitions $A_3$ and $A_4$ are also predicted if there is interplay between two or more control parameters ($n \geq 2$). In this scenario, $\phi_q(t)$ and $\phi_q^S(t)$ can be approximated by the logarithmic expansion

$$\phi_q(t) \sim \left[ f_q - H_q' \ln(t/\tau_\beta) + H_q'' \ln^2(t/\tau_\beta) \right] \quad (1)$$

Together with the prefactors $H_q'$ and $H_q''$, $f_q$ depends both on the wave vector $q$ and on the distance of the state point from the singularity (also known as *separation parameter*, $|x - x_c|$). The characteristic time $\tau$, instead, depends only on the separation parameter and diverges at the transition point. This formula is obtained by asymptotic solution of the MCT equations, assuming that the separation parameter is small (of order $\varepsilon$). It is applicable in the intermediate time range of $\phi_q(t)$, while at longer times it displays the more common $\alpha$-relaxation. This relaxational signature is commonly attributed to a competition between two different arrest mechanisms, usually excluded volume effect and short-range attraction. Doster et al. [28] have applied the $A_2$ formalism of MCT to interpret the quasi-elastic neutron spectrum of protein powder.

In this paper, we show by means of molecular dynamics (MD) simulations that the protein self-intermediate scattering functions display a logarithmic decay in the picosecond to nanosecond time range, that can be fitted according to Eq. 1. In a longer time range, instead, the complete time dependence of the function can be fitted with an analytical model as follows:

$$\phi_q^S(t) \sim \left[ f_q - H_q' \ln(t/\tau^\beta) + H_q'' \ln^2(t/\tau^\beta) \right] \exp\left(-t/\tau_q^\alpha\right) \quad (2)$$

where $\tau^\beta$ and $\tau_q^\alpha$ are the characteristic $\beta$- and $\alpha$-relaxation time, respectively.

We ran MD simulations of a hydrated protein powder model [29] for the lysozyme case. We implemented the OPLS-AA force field [30] for the two lysozyme molecules (PDB file: 1AKI) and the TIP4P-Ew model [31] for the 484 water molecules. Since each protein is composed of 1960 atoms, the total number of atoms in the system was 5872 (including 16 Cl$^-$ ions to neutralize the system) and the triclinic box size was ~ 37 x 42 x 32 Å. After equilibrating the system at 300 K for 50 ns in the *NPT* ensemble ($P$ = 1 bar), we ran 50 ns trajectories at $T$ = 280, 300 and 320 K (i.e. around the physiological temperature, 310 K) in the *NVT* ensemble with a 2 fs timestep. We used a parallel-compiled version of Gromacs 4.0 [32]; we showed in the past that this model correctly reproduces the dynamics of protein hydration water [33]. We also ran one long simulation (500 ns, several months of CPU time) at 310 K to observe the complete long-time decay of the protein self-intermediate scattering functions.

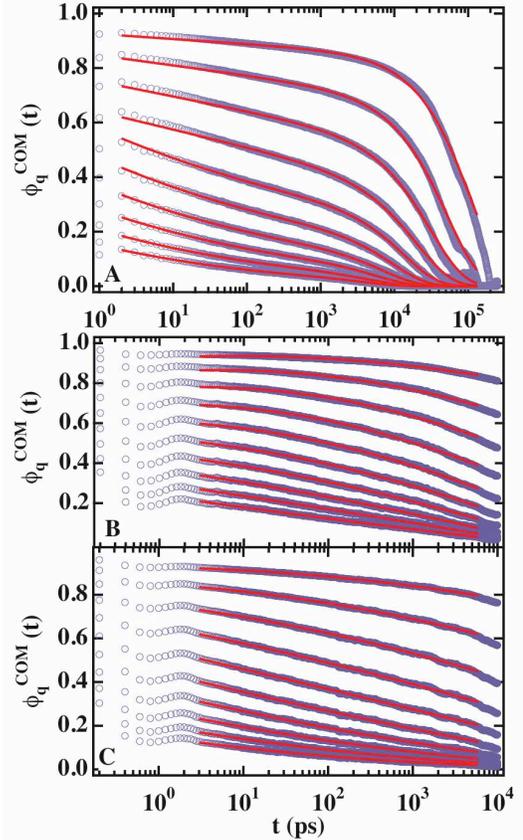

Figure 2. (Color online) Q-vector and temperature dependence of the self-intermediate scattering functions for the COM of the amino acid residues. A) $T$ = 310K, 500ns simulation; B) $T$ = 280, 50ns simulation K; C) $T$ = 320 K, 50ns simulation. 10 different wave vectors are displayed, from 1.6 Å$^{-1}$ to 8.8 Å$^{-1}$ with a 0.8 Å$^{-1}$ interval (from top to bottom). The red continuous lines are the best fits with Eq. 2 (panel A) and Eq. 1 (B and C).

Figure 1 shows the protein-glass analogy in a graphic way: while panel A displays an all-atom representation of a lysozyme molecule, panel B displays only the center-of-mass (COM) of the 129 amino acid residues of the protein. From this representation, it is possible to see how a single-molecule system like a globular native protein could resemble a many-body system like a dense short-ranged attractive colloidal solution. This is quantitatively taken into account in panel D, where we show the liquid-like static structure factor $S(q)$ of the COMs.

Since the partial specific volume of lysozyme is 0.757 cm$^3$/g [34] and the sum of the van der Waals volume of its atoms is 11.8 nm$^3$ [35], we can roughly estimate a volume fraction of $\phi = 0.66$ (close to the value $\phi = 0.61$ used in ref. [17]).

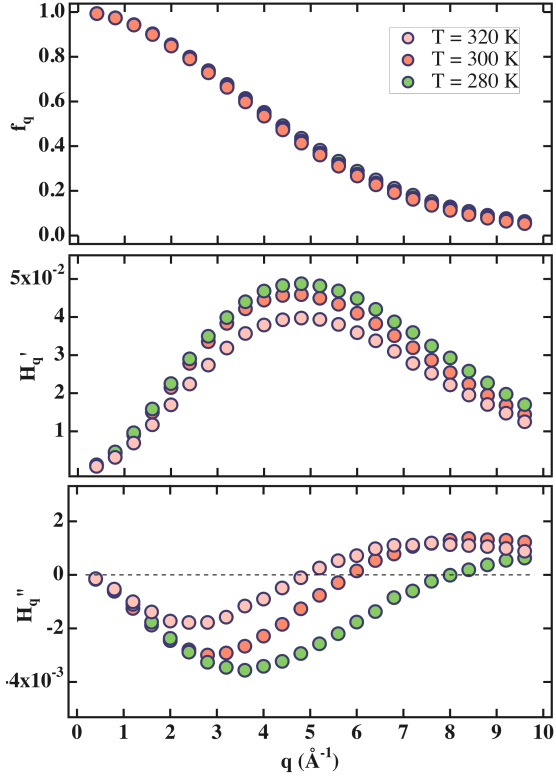

Figure 3. (Color online) Fitting parameters of Eq. 1 for the COM SISF as a function of $q$, for 3 different temperatures. *Upper panel*: Debye-Waller factor, $f_q$. *Middle panel*: first coefficient, $H_q'$. *Bottom panel*: second coefficient $H_q''$.

In Figure 2 we show COM correlators $\phi_q^S(t)$ at various temperatures, for a representative set of wave vectors $q$. It is evident that the relaxation is far from being the classic two-step decay of a liquid-glass transition, and it is not possible to fit the curves with the stretched exponential form we used for protein hydration water [33]. Instead, fitting the correlators with Eq. 2 produces a very good agreement. We would like to point out here that the logarithmic decay is a feature displayed by the $\phi_q^S(t)$ of any kind of atom belonging to the protein (H, C, O…), but considering only the COM of each amino acid residue is the most convenient choice if one wants to exclude the effect of rotations on the correlators.

In Figure 3 we show the $q$ dependence of the fitting parameters of the quadratic polynomial in $\ln(t/\tau)$ reported in Eq. 1, for three different temperatures.

Several predictions of the MCT are verified [17,19]:

1) the Debye-Waller factor $f_q$ (Fig. 3, upper panel) does not depend on the state point, as expected if the system is close to the singularity (correction of order $\varepsilon$ cannot be detected).

2) $H_q'$ can be factorized as $H_q' \equiv h(q)B'(x)$ where $h(q)$ only depends on $q$ and $B'$ only depends on the control parameter $x$. In fact, the $q$-dependence of $H_q'$ is the same for all the temperatures, as displayed in the upper panel of Fig. 4.

3) $H_q''$ does not display the same behavior as $H_q'$, since $B''$ is also a function of $q$. Moreover, $|H_q''| < |H_q'|$ since the first is of order $\varepsilon$ and the second of order $\varepsilon^{1/2}$.

4) the $q$-values where $H_q'' = 0$ border a convex-to-concave crossover, as predicted by the theory. This is one of the main signatures of the higher-order MCT scenario. These $q$-values depend on the state point.

5) The correlators collapse on the logarithmic decay law $-\ln(t/\tau)$ if they are rescaled as $(\phi_q(t) - f_q)/H_q'$ (Fig. 4 lower panel).

The $q$-dependence of the protein $\alpha$-relaxation time extracted from Eq. 2 at $T = 310$ K is shown in Fig. 5. We find that $1/\tau_q^\alpha \sim Dq^2$, with the diffusion constant $D = 3.1 \cdot 10^{-10}$ cm$^2$/s, indicating a glassy liquid-like diffusive behavior for the constituents of the protein at physiological temperatures. Compared to a common glass-forming liquid like o-terphenyl, that also shows a logarithmic decay [10], this magnitude of the diffusion constant would correspond to $T \sim 290$ K, around the crossover temperature $T_c$ [36].

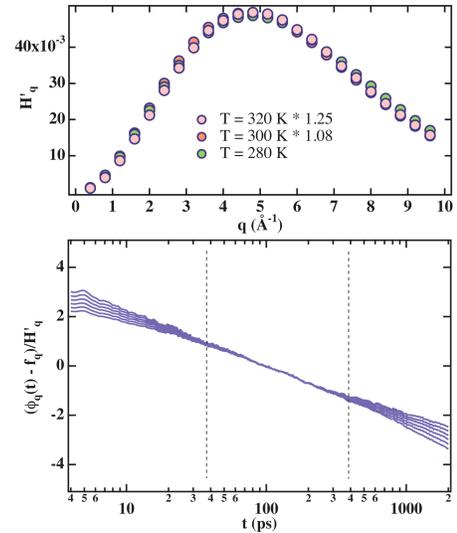

Figure 4. (Color online) Scaling plots of the MCT predictions. *Upper panel*: all the $H_q'$ at different temperatures collapse on top of each other if multiplied by proper factors. *Lower panel*: Scaled tagged-particle correlation functions at $T = 300$ K for $q = 3.2, 4.0, 4.8, 5.6, 6.4, 7.2$ Å$^{-1}$. The values of $\tau$ are 25, 100 and 600 ps for $T = 320, 300$ and 280 K respectively. The vertical dashed lines indicate the time interval where the first order approximation holds.

In conclusion, we showed a logarithmic decay of the protein tagged-particle correlators by means of MD

simulations. This anomalous behavior resembles the MCT results for dense liquids close to a higher-order glass transition, and suggests that a globular protein can be seen as a close-packed colloidal system. In particular, the complete decay of the ISF to zero at the physiological temperature is further proof that the functioning proteins behave like a glassy liquid [37,38], and not like a solid.

We would like to stress here that the agreement between the protein dynamics and the predictions of the idealized MCT does not provide evidence of the existence of a higher-order singularity in proteins. Only solving the MCT equations for this hetero-polymeric system could provide an appropriate answer, and at present the MCT equations have only been solved for homo-polymers [39]. Nevertheless, mapping the protein dynamics onto the dynamics of a short-ranged attractive colloidal system reinforces the analogy between globular proteins and glass-forming liquids, and adds a piece to the puzzle of the interplay between the dynamics and the biological function of biomolecules.

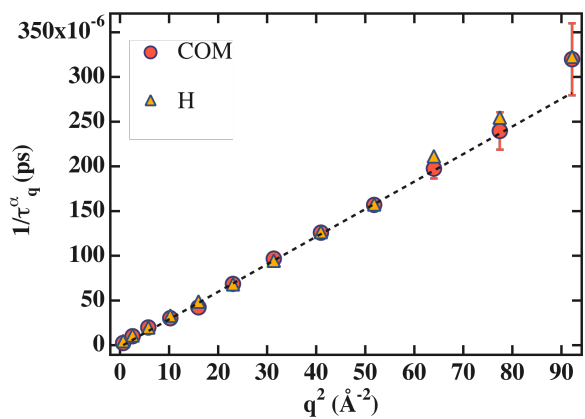

Figure 5. (Color online) $q$-dependence of the $\alpha$-relaxation time extracted from Eq. 2 for hydrogens and COM, at T = 310 K.

The next important question to be addressed is in fact why Nature has chosen such a relaxational behavior for proteins. A possible answer could be that the logarithmic decay is the slowest possible time dependence of motion, and this could endow proteins with the appropriate resilience in response to the fluctuations of the external environment.

Research at MIT is supported by DOE Grants No. DE-FG02-90ER45429. M.L. and P.B. acknowledge financial support from CSGI and MIUR. We benefited from affiliation with EU funded Marie-Curie Research and Training Network on Arrested Matter. We thank Francesco Sciortino and Emanuela Zaccarelli for helpful discussions.